\documentclass[twocolumn]{aastex631}

\newcommand{\red}{\textcolor{black}}

\begin{document}

\title{Detection of extended X-ray emission around the PeVatron microquasar V4641 Sgr with XRISM}

\author[0000-0002-8152-6172]{Hiromasa~Suzuki}
\affiliation{Institute of Space and Astronautical Science (ISAS), Japan Aerospace Exploration Agency (JAXA), Kanagawa 252-5210, Japan}
\author[0000-0001-7209-9204]{Naomi Tsuji}
\affiliation{Faculty of Science, Kanagawa University, 3-27-1 Rokukakubashi, Kanagawa-ku, Yokohama-shi, Kanagawa 221-8686, Japan}
\affiliation{Interdisciplinary Theoretical \& Mathematical Science Program (iTHEMS), RIKEN, 2-1 Hirosawa, Wako, Saitama 351-0198, Japan}
\author[0000-0002-4541-1044]{Yoshiaki Kanemaru}
\affiliation{Institute of Space and Astronautical Science (ISAS), Japan Aerospace Exploration Agency (JAXA), Kanagawa 252-5210, Japan}
\author[0000-0001-8195-6546]{Megumi Shidatsu}
\affiliation{Department of Physics, Ehime University, 2-5 Bunkyocho, Matsuyama, Ehime 790-8577, Japan}
\author[0000-0002-9105-0518]{Laura Olivera-Nieto}
\affiliation{Max-Planck-Institut f\"{u}r Kernphysik, P.O. Box 103980, D 69029 Heidelberg, Germany}
\author[0000-0001-6189-7665]{Samar Safi-Harb}
\affiliation{Department of Physics and Astronomy, University of Manitoba, Winnipeg, MB R3T 2N2, Canada}
\author[0000-0003-2579-7266]{Shigeo S. Kimura}
\affiliation{Frontier Research Institute for Interdisciplinary Sciences, Tohoku University, Sendai 980-8578, Japan}
\affiliation{Astronomical Institute, Graduate School of Science, Tohoku University, Sendai 980-8578, Japan}
\author[0000-0001-9643-4134]{Eduardo de la Fuente}
\affiliation{Departamento de F\'{i}sica, CUCEI, Universidad de Guadalajara, Blvd. Marcelino Garc\'{i}a Barragan 1420, Ol\'{i}mpica, 44430, Guadalajara, Jalisco, M\'exico}
\author[0000-0002-6144-9122]{Sabrina Casanova}
\affiliation{Instytut Fizyki J\c{a}drowej PAN, ul. Radzikowskiego 152, 31-342 Krak{\'o}w, Poland}
\author[0000-0002-9709-5389]{Kaya Mori}
\affiliation{Columbia Astrophysics Laboratory, Columbia University, 538 West 120th Street, New York, NY 10027, USA}
\author[0000-0001-6798-353X]{Xiaojie Wang}
\affiliation{Department of Physics, Michigan Technological University, Houghton, MI, USA}
\author{Sei Kato}
\affiliation{Institute for Cosmic Ray Research, University of Tokyo, Kashiwa 277-8582, Japan}
\affiliation{Institut d'Astrophysique de Paris, CNRS UMR 7095, Sorbonne Universite, 98 bis bd Arago 75014, Paris, France (Present affiliation)}
\author[0000-0003-0248-4064]{Dai Tateishi}
\affiliation{Department of Physics, Graduate School of Science, The University of Tokyo, 7-3-1 Hongo, Bunkyo-ku, Tokyo 113-0033, Japan}
\author[0000-0003-4580-4021]{Hideki Uchiyama}
\affiliation{Faculty of Education, Shizuoka University, 836 Ohya, Suruga-ku, Shizuoka, Shizuoka, 422-8529, Japan}
\author[0000-0002-4383-0368]{Takaaki Tanaka}
\affiliation{Konan University, Department of Physics, 8-9-1 Okamoto, Higashinada, Kobe, Hyogo, Japan, 658-8501}
\author[0000-0003-1518-2188]{Hiroyuki Uchida}
\affiliation{Department of Physics, Graduate School of Science, Kyoto University, Kitashirakawa Oiwake-cho, Sakyo-ku, Kyoto 606-8502, Japan}
\author[0000-0003-3085-304X]{Shun Inoue}
\affiliation{Department of Physics, Graduate School of Science, Kyoto University, Kitashirakawa Oiwake-cho, Sakyo-ku, Kyoto 606-8502, Japan}
\author[0000-0002-5447-1786]{Dezhi Huang}
\affiliation{Department of Physics, University of Maryland, College Park, MD, USA}
\author[0000-0002-4462-3686]{Marianne Lemoine-Goumard}
\affiliation{Universite Bordeaux, CNRS, CENBG, UMR 5797, F-33170 Gradignan, France}
\author[0009-0009-0439-1866]{Daiki Miura}
\affiliation{Department of Physics, Graduate School of Science, The University of Tokyo, 7-3-1 Hongo, Bunkyo-ku, Tokyo 113-0033, Japan}
\affiliation{Institute of Space and Astronautical Science (ISAS), Japan Aerospace Exploration Agency (JAXA), Kanagawa 252-5210, Japan}
\author[0000-0002-5701-0811]{Shoji Ogawa}
\affiliation{Institute of Space and Astronautical Science (ISAS), Japan Aerospace Exploration Agency (JAXA), Kanagawa 252-5210, Japan}
\author[0000-0001-7773-9266]{Shogo B. Kobayashi}
\affiliation{Department of Physics, Tokyo University of Science, 1-3 Kagurazaka, Shinjuku-ku, Tokyo 162-8601, Japan}
\author[0000-0002-1065-7239]{Chris Done}
\affiliation{Centre for Extragalactic Astronomy, Department of Physics, University of Durham, South Road, Durham DH1 3LE, UK}
\author[0009-0003-8610-853X]{Maxime Parra}
\affiliation{Universite Grenoble Alpes, CNRS, IPAG, 38000 Grenoble, France}
\affiliation{Dipartimento di Matematica e Fisica, Universita degli Studi Roma
Tre, Via della Vasca Navale 84, 00146 Roma, Italy}
\author[0000-0001-7796-4279]{Maria D\'iaz Trigo}
\affiliation{ESO, Karl-Schwarzschild-Strasse 2, 85748, Garching bei M\"unchen, Germany}
\author[0000-0002-3348-4035]{Teo Mu$\tilde{\rm n}$oz-Darias}
\affiliation{Instituto de Astrofisica de Canarias, E-38205 La Laguna, Tenerife, Spain}
\affiliation{Departamento de Astrofisica, Universidad de La Laguna, E-38206 La Laguna, Tenerife, Spain}
\author[0000-0002-4344-7334]{Montserrat Armas Padilla}
\affiliation{Instituto de Astrofisica de Canarias, E-38205 La Laguna, Tenerife, Spain}
\affiliation{Departamento de Astrofisica, Universidad de La Laguna, E-38206 La Laguna, Tenerife, Spain}
\author[0000-0002-6797-2539]{Ryota Tomaru}
\affiliation{Department of Earth and Space Science, Graduate School of Science, Osaka University, 1-1 Machikaneyama, Toyonaka, Osaka 560-0043, Japan}
\author{Yoshihiro Ueda}
\affiliation{Department of Astronomy, Kyoto University, Kitashirakawa-Oiwake-cho, Sakyo-ku, Kyoto, Kyoto 606-8502, Japan}

% \author{more...}



\begin{abstract}
% \red{(UNDER REVISION)} 
A recent report on the detection of very-high-energy gamma rays from V4641 Sagittarii (V4641 Sgr) up to $\approx 0.8$~peta-electronvolt has made it the second confirmed ``PeVatron'' microquasar.
% The nature of its gamma-ray emission is unclear because X-ray and radio properties of the gamma-ray emitting region have been completely unknown.
Here we report on the observation of V4641 Sgr with X-Ray Imaging and Spectroscopy Mission (XRISM) in September 2024. Thanks to the large field of view and low background, the CCD imager Xtend successfully detected for the first time X-ray extended emission around V4641 Sgr with a significance of %$\gtrsim 4.5\sigma$ from a morphological analysis and $>10\sigma$ from spectroscopy. 
$\gtrsim 4.5\sigma$ and $>10\sigma$ based on our imaging and spectral analysis, respectively.
The spatial extent is estimated to have a radius of $7\pm3$~arcmin ($13\pm5$~pc at a distance of 6.2~kpc) assuming a Gaussian-like radial distribution, which suggests that the particle acceleration site is within $\sim 10$~pc of the microquasar. If the X-ray morphology traces the diffusion of accelerated electrons, this spatial extent can be explained by either an enhanced magnetic field ($\sim 80~\mu$G) or a suppressed diffusion coefficient ($\sim 10^{27}$~cm$^2$~s$^{-1}$ at 100~TeV).
%
% We found that the spectral shape can be explained by either a power-law (non-thermal) or an optically-thin thermal plasma model.
The integrated X-ray flux, (4--$6)\times10^{-12}$~erg~s$^{-1}$~cm$^{-2}$ (2--10 keV), would require a magnetic field strength higher than the galactic mean ($\gtrsim 8~\mu$G) 
%if the X-rays are of non-thermal origin and the gamma-rays are predominantly hadronic.
if the diffuse X-ray emission originates from synchrotron radiation and the gamma-ray emission is predominantly hadronic.
If the X-rays are of thermal origin, the measured extension, temperature, and plasma density can be explained by a jet with a luminosity of $\sim 2\times10^{39}$~erg~s$^{-1}$, which is comparable to the Eddington luminosity of this system. 

% is higher than that expected from a one-zone hadronic model with the reported gamma-ray flux.
% The magnetic field strength in a one-zone leptonic model is constrained to be 3--7~$\mu$G.
% A natural explanation is that the detected X-rays originate from an electron population near the acceleration sites close to V4641 Sgr, which is different than those responsible for the gamma-ray emission.

\end{abstract}

\keywords{Low-mass x-ray binary stars (939) --- Gamma-ray sources (633) --- Radio jets (1347) --- Non-thermal radiation sources (1119)
}


\section{Introduction} \label{sec-intro}
The origin of Galactic cosmic rays has been a long-standing question and is of increasing interest in recent years thanks to the growing high-sensitivity X-ray, gamma-ray, and neutrino observatories.
Among the candidates for Galactic cosmic-ray accelerators, microquasars, where accreting black holes are powering jets,
%are rapidly gaining attention in these years, following the tera-electronvolt (TeV) detection of the W50-SS433 system \citep{hawc18_ss433, hess24_ss433}.
are an emerging class of Galactic PeVatrons after the tera-electronvolt (TeV) detection of the W50-SS~433 system \citep{hawc18_ss433, hess24_ss433}.
Most recently, LHAASO discovered gamma-ray emission exceeding 100~TeV from directions overlapping with several microquasars \citep{lhaaso24_v4641}.
The detection of gamma rays with $E>100$~TeV suggests that parent particles (either protons or electrons) are accelerated beyond 1~PeV, indicating these microquasars are potential PeVatrons.

%V4641 Sagittarii (V4641 Sgr), the brightest source in the TeV-emitting microsuasars, was found to exhibit extended gamma-ray emission with the photon energies up to $\approx 0.8$~peta-electronvolt (PeV) \citep{hawc24_v4641, lhaaso24_v4641}.
V4641 Sagittarii (V4641 Sgr) was found to exhibit extended gamma-ray emission up to $\approx 0.8$~peta-electronvolt (PeV) in its surroundings, which is the brightest among the TeV-detected microquasars \citep{hawc24_v4641, lhaaso24_v4641}.
% These discoveries have revived interest in microquasars as potential accelerators of Galactic cosmic rays. 
% Microquasars, as represented by SS433, are often discovered as intense X-ray binaries and are considered an important class of galactic particle accelerators \citep{hawc18_ss433, hess24_ss433}. 
V4641 Sgr is a low-mass X-ray binary (LMXB) consisting of a black hole with a mass of $6.4 \pm 0.6\, M_{\odot}$ and a B9III companion star with a mass of $2.9 \pm 0.4\, M_{\odot}$ \citep{macdonald14}, located at a distance of $\approx 6.2$~kpc \citep{macdonald14, gandhi19}.
In contrast to other microquasars, V4641 Sgr is characterized by violent X-ray outbursts, which include a rapid onset and exponential decay \citep{revnivtsev02}. The outbursts are observed once in $\sim 2$~years without clear periodicity\footnote{Refer to, e.g., a light curve by MAXI (\url{http://maxi.riken.jp/star_data/J1819-254/J1819-254.html}).}.
%The outburst rate is estimated to be once in $\sim 220$~days \cite{tetarenko16}.
Radio observations with the Very Large Array, performed within one day after an X-ray burst in 1999, showed a luminous jet-like radio structure of about $0\farcs25$ length \citep{hjellming00}, much more compact than the gamma-ray extension of $\approx 0\fdg54$ \citep{hawc24_v4641}.

\begin{figure}
    \centering
    \includegraphics[width=8cm]{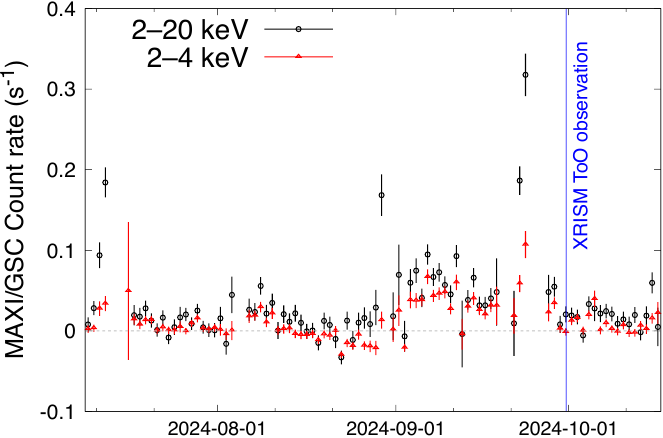}
    \caption{\red{MAXI/GSC light curves of V4641 Sgr showing the bursting activity in September 2024, obtained from \url{http://maxi.riken.jp/star_data/J1819-254/J1819-254.html}. The observation period with XRISM is indicated with the blue transparent region.}}
    \label{fig-maxi}
\end{figure}

In these months, V4641 Sgr is gaining even more attention because it has been exhibiting a bursting activity since September 2024 \red{(Figure~\ref{fig-maxi}; report on MAXI/GSC observations: \url{https://www.astronomerstelegram.org/?read=16804})}. Following other X-ray observatories such as Swift and XMM-Newton, the X-Ray Imaging and Spectroscopy Mission (XRISM; \citealt{tashiro18, tashiro20, tashiro24}) observed V4641 Sgr as a generic Target of Opportunity (ToO) observation.
In this paper, we report on a morphological and spectral analysis of the environment of V4641 Sgr and the detection of extended emission with XRISM.

\section{Observation and data reduction} \label{sec-obs}

\begin{figure*}[htb!]
    \centering
    \includegraphics[width=16cm]{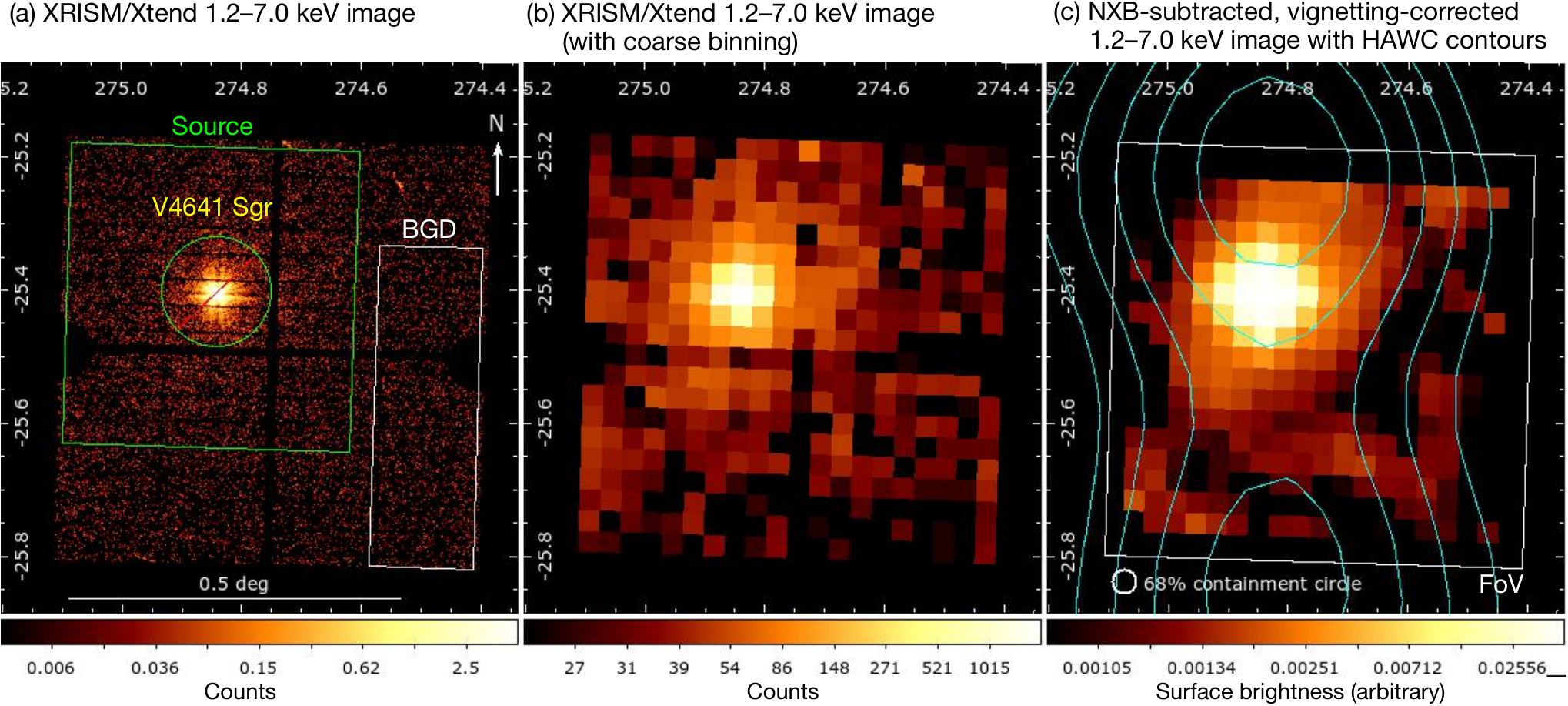}
    \caption{(a) A 1.2--7.0~keV XRISM Xtend image with the regions for the spectral analysis. (b) Same with a course binning. (c) The image after the particle-background subtraction and vignetting correction and with the HAWC gamma-ray contours ($>1$~TeV; \citealt{hawc24_v4641}) and the Xtend field of view (FoV) overlaid. The 68\% containment circle of the point spread function at 6.4~keV is shown as a reference. The images are shown on the Equatorial coordinates (J2000).}
    \label{fig-image}
\end{figure*}

XRISM is equipped with the microcalorimeter Resolve \citep{ishisaki18} and wide field-of-view (FoV) CCD camera Xtend \citep{mori22}. In this work, we focus on the data obtained with Xtend to search for a possible extended emission.\footnote{The results of the high-resolution spectroscopy of V4641 Sgr itself with Resolve will be reported separately (Shidatsu~et~al., in prep.).}
% Resolve consists of 35 active pixels, covering a sky region $\approx 0\farcm5 \times 0\farcm5$ each and $\approx 3\farcm1 \times 3\farcm1$ combined, with an energy resolution of $< 5$~eV (FWHM: full width half maximum; \cite{xrism24_n132d}).
Xtend provides a $\approx 38'\times38'$ FoV for the energy range 0.4--13~keV with an energy resolution better than 200~eV (FWHM: full width half maximum; \citealt{mori24}). The angular resolution is $< 1\farcm7$ (HPD: half power diameter; \citealt{tamura24}).

V4641 Sgr was observed with XRISM as a generic ToO target on September 30, 2024 UT (ObsID: 901001010). Since the allocated time for generic ToO observations is very limited, the observation duration was only $\approx 20$~ks.
% Resolve provides the desired energy resolution of 4.4~eV (FWHM), confirmed based on the Mn-K$\alpha$ spectrum from the $^{55}$Fe onboard radioactive sources.
% We processed data with the current standard method recommended by the XRISM team.
We reduced the data with the pre-release Build 7 XRISM software with HEAsoft ver. 6.32 \citep{heasarc14} and calibration database (CALDB) ver. 8 (v20240815) \citep{terada21, loewen20}.
% The data are processed and screened by the automated XRISM processing pipeline ver. 03.00.011.008.
We excluded periods of the Earth eclipse and sunlit Earth's limb, as well as passages of the South Atlantic Anomaly.
% More details of the data reduction are described in \cite{xrism24_n132d}.
The effective exposure of Xtend that remains after the standard data reduction is $\approx 12.2$~ks. 
% We note that the Resolve observations were made through a $\sim 250~\mu$m thick beryllium window \cite{midooka20} in the closed aperture door, limiting the bandpass to energies above $\sim1.6$~keV.
% The systematic uncertainty in the energy scale of Resolve is evaluated using the onboard $^{55}$Fe source and found to be very small, $\approx 0.1$~eV at 5.9~keV {\citep{eckart24, porter24}.}
% This translates to a velocity of $\approx 5$~km~s$^{-1}$.
We removed flickering pixels\footnote{Pixels with anomalously high event rates, which mostly record pseudo events at $\lesssim 1$~keV \citep{nakajima18}.} using the {\tt searchflickpix} tool in addition to removing bad pixels and columns manually.

For the morphological analysis, we use the night-earth and day-earth occultation data collected from March 10 to July 31, 2024 from the XRISM trend archive (Rev.~3).\footnote{\url{https://data.darts.isas.jaxa.jp/pub/xrism/data/trend/rev3/}}
The night-earth data are used to evaluate the event rate and spatial distribution of the particle background (Non-X-ray background), whereas the day-earth data are used as a reference for the detector image from a uniform emission on the plane of the sky, which allows us to obtain the vignetting curve (effective area as a function of the off-axis angle).

In our spectral analysis, we use XSPEC ver. 12.13.1 \citep{arnaud96} with the $C$-statistic \citep{cash79}, and AtomDB ver. 3.0.9 \citep{foster12}.
% In calculation of charge exchange X-ray emission, we partly use the CX model \citep{gu16} in SPEX ver. 3.08.00 \citep{kaastra96}.
Redistribution Matrix Files (RMFs) are created with the {\tt xtdrmf} task using the cleaned event file and CALDB based on ground measurements.
% Line-spread function components include the Gaussian core, exponential tail to low energy, escape peaks, and Si fluorescence.
Auxiliary Response Files (ARFs) are generated with the {\tt xaarfgen} task assuming constant surface-brightness emission because the sky background dominates the data.
% The assumed emission profiles for individual spectral components are described in Section~\ref{sec-spec}. 
The errors given in the text and figures indicate $1\sigma$ confidence intervals.

\section{Analysis and results} \label{sec-ana}

\subsection{X-ray morphology}\label{sec-image}
Figure~\ref{fig-image} (a) and (b) show a 1.2--7.0~keV image with XRISM Xtend, indicating a hint of extended emission around V4641 Sgr. Here, events with lower and higher energies are excluded because of the significant contribution of the sky and particle background, respectively (see Section~\ref{sec-spec}).
To be more precise, we subtract the particle-background image extracted from the night-earth dataset and scaled to match the 9.0--13.0~keV event rate, dominated by the particle background, of the V4641 Sgr observation. Then we also correct for the vignetting of the mirror assembly using the day-earth image.
As a result, we obtain Figure~\ref{fig-image} (c), which shows extended emission around V4641 Sgr.

\begin{figure*}[htb!]
    \centering
    \includegraphics[width=16cm]{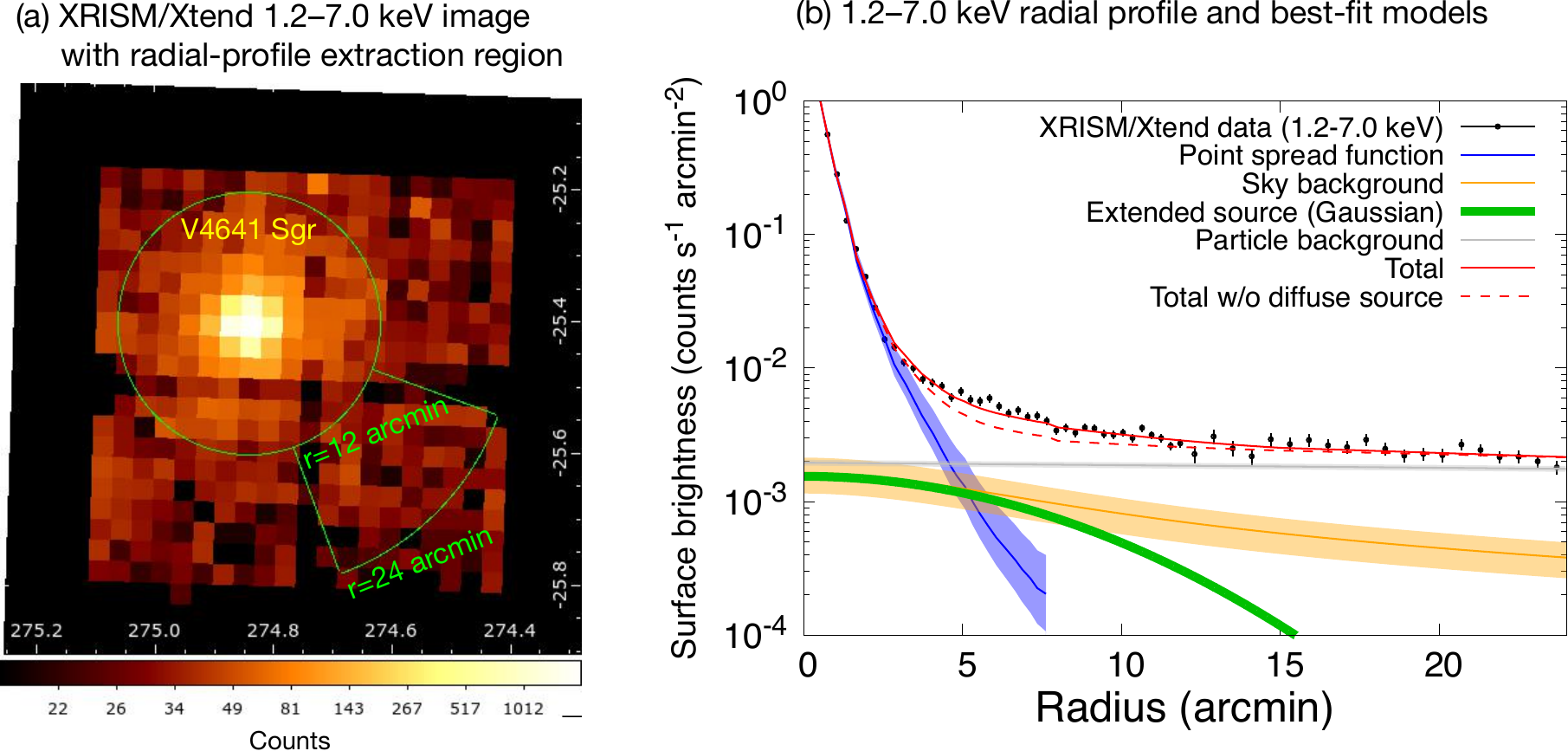}
    \caption{(a) A 1.2--7.0~keV XRISM Xtend image with the radial-profile extraction region centered at V4641 Sgr (green). (b) Extracted radial profile with the models for the point spread function at 6.4~keV, sky background, particle background, and extended emission modeled with a Gaussian function. The transparent regions indicate the systematic uncertainty ranges (see text for details).}
    \label{fig-radial}
\end{figure*}

In order to quantify the spatial extent of the extended emission, we extract a one-dimensional radial profile centered at V4641 Sgr (Figure~\ref{fig-radial}). 
The profile is extracted from a circle with a radius of 12~arcmin and from a partial annulus extending from 12 to 24~arcmin (Figure~\ref{fig-radial} (a)). The location of the partial annulus is chosen to extend to the region where events are dominated by particle background.
The radial profile for the emission of V4641 Sgr (point spread function: PSF) is made with a ray-tracing simulation using the {\tt xrtraytrace} task at 6--7~keV\footnote{The in-orbit calibration of the PSF has been done best for this energy range. The energy dependence of the PSF in 1.2--7.0~keV ($< 80\%$ at radii smaller than 8~arcmin: \citealt{tamura22}) is included in the systematic uncertainty considered later.}. The morphology of the sky background, which should originate from a nearly uniform emission on the plane of the sky, is modeled using the day-earth image, which is equivalent to the vignetting image. The radial profile of the particle-background is extracted from the night-earth dataset.
As for the the extended emission, we assume a Gaussian-like distribution as a simple model. We model the radial profile from the 1.2--7.0~keV image with the empirical model ``(V4641 Sgr) + (extended emission) + (sky background) + (particle background)''. The free parameters include the normalizations of the components of V4641 Sgr, the extended emission, and sky background, and the width of the extended emission (Gaussian function).
\red{We employ a least chi-square fit to determine the best-fit parameters and their confidence intervals.} 
The best-fit model is shown in Figure~\ref{fig-radial} (b). We obtain the best-fit \red{Gaussian sigma} of the extended-emission model of $7\pm3$~arcmin, which is converted to $13\pm5$~pc at a distance of 6.2~kpc. The model without the extended emission does not fit the data in the radii of $\approx 5$--12~arcmin (Figure~\ref{fig-radial} (b)). The detection significance can be evaluated from the fit statistics with and without the extended-emission model using the F-test, to be $\gtrsim 4.5\sigma$ (F-test probability $\lesssim 8.8\times10^{-6}$) taking into account the \red{upper limits of the} systematic uncertainties described below.

We evaluate the impact of the systematic uncertainties associated with the PSF tail: $< 80\%$ within the radii $< 8$~arcmin \citep{tamura22, tamura24}, off-axis effective area: $< 30\%$, and particle background level: $<5\%$ (Uchida et al., in prep.). These uncertainty levels are also shown in Figure~\ref{fig-radial} (b). \red{These systematic uncertainties can alter the flux of the emission up to $\approx 30\%$.}

\subsection{X-ray spectra}\label{sec-spec}

\begin{figure*}[htb!]
    \centering
    \includegraphics[width=16cm]{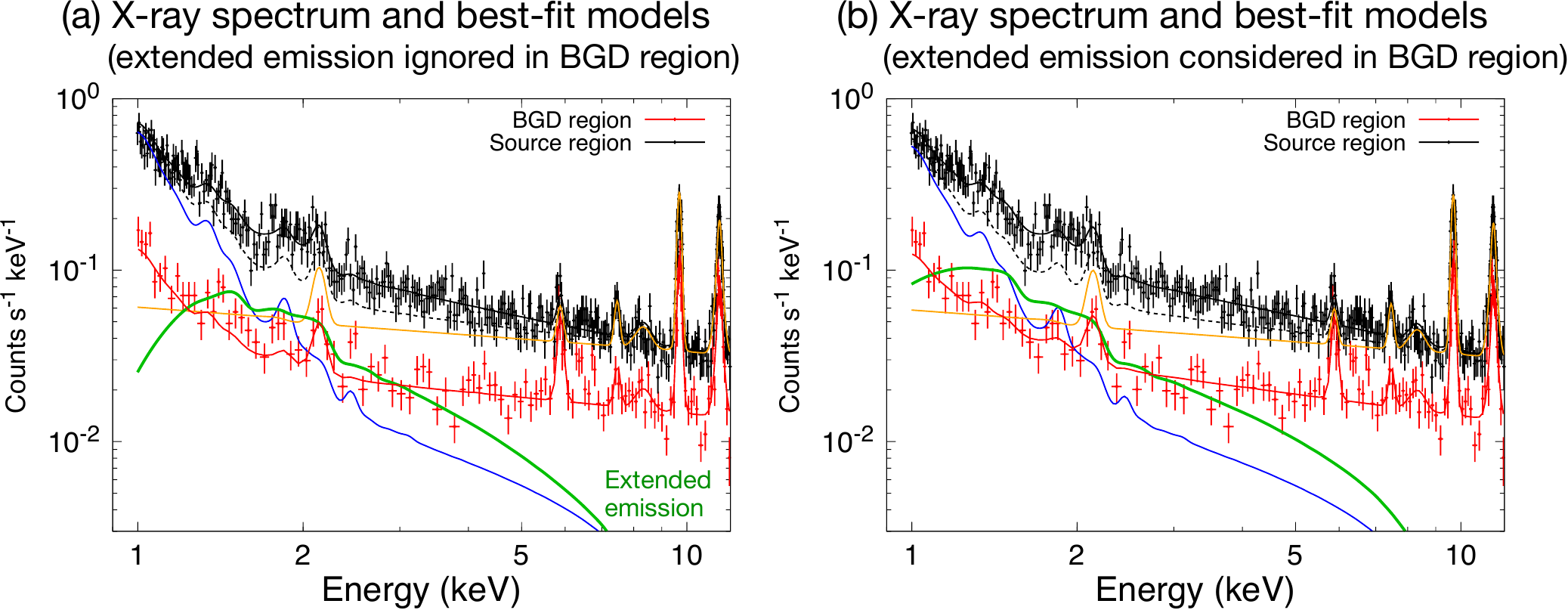}
    \caption{(a) X-ray spectra obtained with XRISM Xtend and the best-fit spectral models in the case where contamination of the extended emission in the BGD region is ignored and (b) assumed to be 10\% in flux of the source region. The green, blue, and orange solid lines indicate the extended emission, sky background, and particle background, respectively, for the source region. The black and red solid lines indicate the total models for the source and BGD regions, respectively. The black dashed line is the total model for the source region but without the extended-emission component.}
    \label{fig-spec}
\end{figure*}

In the next step, we examine the X-ray spectrum of the extended emission \red{marginally detected in Section~\ref{sec-image}}. The source and background (BGD) regions are defined as shown in Figure~\ref{fig-image} (a). The source region encloses most of the \red{apparent} extended emission (Section~\ref{sec-image}) but without the contribution of V4641 Sgr itself\footnote{$\lesssim 3\%$ of the total flux remains within the source region \citep{tamura24}, which is lower than the sky background level.}. The BGD region is nearly free from the extended emission (Figure~\ref{fig-image} (c)).
We only use the energies above 1.0~keV, where the contamination from potential flickering pixels and bad pixels remaining after data reduction is negligible. 

For the spectral model, we consider the sky and particle background in addition to the extended source.
As for the sky background, we apply the foreground emission (Local Hot Bubble), Milky Way Halo (Transabsorption emission) and Cosmic X-ray Background (CXB).
The former two are modeled with the {\tt apec} model in XSPEC with the electron temperatures of 0.1~keV and 0.6~keV, respectively, and the solar abundance \citep{kuntz08, masui09, yeung24}. The uncertain Galactic absorption for the Milky Way Halo component is not considered because it does not strongly affect the energy range we focus on. Only their fluxes are treated as free parameters. As for CXB, we use the {\tt powerlaw} model with a photon index of 1.41 and the 2--10~keV flux fixed to $5.4\times10^{-15}$~erg~cm$^{-2}$~arcmin$^{-2}$ \citep{kushino02}. The Galactic absorption is considered with the {\tt tbabs} model with a column density of $1.6\times10^{21}$~cm$^{-2}$ \citep{wilms00,hi4pi16}. 

The particle background model is developed from the night-earth observation data. The extracted spectrum can be explained with a model composed of the fluorescence lines of Au-M, Ni-K, and Au-L and a continuum ({\tt powerlaw}) (\citealt{nakajima18}; Uchida et al. in prep.). As the particle background level depends on the satellite position and solar activity but without changing the spectral shape significantly ($\lesssim 10\%$; Uchida et al. in prep.), we apply this spectral model to the V4641 Sgr observation. Only the total normalization is treated as a free parameter.

As for the extended emission, we first use an absorbed power-law model ({\tt tbabs$\times$powerlaw}). % assuming a non-thermal nature.
In the spectral modeling, we fit the spectra of the source and BGD simultaneously with two assumptions: (a) the extended emission does not contaminate the BGD region and (b) 10\% flux of the extended emission in the source region contributes to the BGD region. The contamination level of 10\% is \red{derived by evaluating the integrals of a Gaussian function for the extended source, assuming a sigma of 10~arcmin (best-fit value plus 1$\sigma$ error; see Section~\ref{sec-image}) over the source and BGD regions.}
In the spectral modeling, we find that the detector gain is slightly shifted from that expected by the RMFs due to the degradation on orbit in $\sim 1$~year. We thus correct the gain slopes of the RMFs by $\approx 0.3\%$ and $\approx 0.2\%$ for the source and BGD regions, respectively, to fit the detector background lines.

Figure~\ref{fig-spec} shows the results corresponding to the two cases (a) and (b). We obtain the fit statistics $C$-stat/d.o.f. = 4190.1/3658 and 4147.6/3658, respectively. The case (b) better explains the observations.
The best-fit spectral parameters of the extended emission are summarized in Table~\ref{tab-spec}. The unabsorbed flux in 2--10~keV is determined to be $\approx 7\text{--}10 \times 10^{-15}$~erg~s$^{-1}$~cm$^{-2}$~arcmin$^{-2}$. The spectral index is found to depend on the assumptions strongly, and thus is not constrained well.
The detection significance of the extended emission can be evaluated from the fit statistics with and without the extend-emission model using the likelihood ratio test (for the $C$-statistic), to be $> 10\sigma$ ($\Delta C = 361.9$ and $\Delta$d.o.f. = 3).

To account for the possibility that the emission is of thermal origin, we also assume a thermal emission model (case (c)). If a collisional ionization equilibrium plasma model ({\tt apec}) with the solar abundance \citep{wilms00} and with free electron temperature and normalization replaces the power-law model of the case (a), we obtain the fit statistics $C$-stat/d.o.f. = 4192.3/3658. This is similar to the result of the case (a). The derived parameters are shown in Table~\ref{tab-spec}. The electron temperature $kT_{\rm e}$ is constrained to be $3.2\pm0.7$~keV.
% , which is higher than that observed in SS433 \citep{kayama22}. 
The unabsorbed flux in 2--10~keV is estimated to be $(7\pm1) \times 10^{-15}$~erg~s$^{-1}$~cm$^{-2}$~arcmin$^{-2}$.
Based on the volume emission measure $n_{\rm e}n_{\rm p}V$, where the $n_{\rm e}$, $n_{\rm p}$, and $V$ respectively are the electron and proton density, and volume, one can derive the plasma density $\approx 0.3~{\rm cm}^{-3} \,(r/10~{\rm pc})^{-1.5}$, with the $r$ being the radius of the spherical plasma.

\begin{table}[htb!]
    \centering
    \caption{Best-fit spectral parameters of the extended emission.}
    \begin{tabular}{l l c}
    \hline\hline
        Case (a) & $N_{\rm H}$ ($10^{22}$~cm$^{-2}$) & $1.8\pm0.5$ \\
         & Power-law index & $2.4\pm0.3$  \\
         & Power-law flux (2--10~keV) & $7.8\pm0.9$  \\
         & ($10^{-15}$~erg~s$^{-1}$~cm$^{-2}$~arcmin$^{-2}$) &   \\
         & Integrated power-law flux (2--10~keV)\tablenotemark{a} & $4.3\pm0.5$  \\
         & ($10^{-12}$~erg~s$^{-1}$~cm$^{-2}$) &   \\
         & $C$-stat./d.o.f. & 4190.1/3658  \\
        Case (b) & $N_{\rm H}$ ($10^{22}$~cm$^{-2}$) & $0.6\pm0.4$ \\
         & Power-law index & $1.8\pm0.2$  \\
         & Power-law flux (2--10~keV) & $9.2\pm1.1$  \\
         & ($10^{-15}$~erg~s$^{-1}$~cm$^{-2}$~arcmin$^{-2}$) &   \\
         & Integrated power-law flux (2--10~keV)\tablenotemark{a} & $5.1\pm0.6$  \\
         & ($10^{-12}$~erg~s$^{-1}$~cm$^{-2}$) &   \\
         & $C$-stat./d.o.f. & 4147.6/3658  \\
        Case (c) & $N_{\rm H}$ ($10^{22}$~cm$^{-2}$) & $1.5\pm0.4$ \\
         & $kT_{\rm e}$ (keV) & $3.2\pm0.7$ \\
         & $n_{\rm e}n_{\rm p}V$ (10$^{58}$~cm$^{-3}$) & $1.3\pm0.2$ \\
         & $C$-stat./d.o.f. & 4192.3/3658  \\
    \hline
    \end{tabular}
    \label{tab-spec}
    \tablenotetext{a}{Integrated flux for the source region, which has an area of 629.4~arcmin$^2$.}
\end{table}

Considering its location ($\ell=6\fdg77$\ and $b=-4\fdg79$) close to the Galactic center, the spectrum is possibly affected by the Galactic Ridge X-ray Emission (GRXE:  \citealt{uchiyama13, yamauchi16}), but the exact contribution at such high galactic altitudes is quite uncertain.
So, we add a GRXE model (with the spectral shape described in Table~3 of \citealt{uchiyama13}) with the highest flux level allowed from the best-fit radial-profile models (Section~\ref{sec-image})\footnote{Total sky-background rate of $\approx 70\%$ particle-background rate in 1.2--7.0~keV for radii of 5--15~arcmin, roughly corresponding to the source region.}. As a result, the spectral excess is still found to be significant with a $\approx 9\sigma$ significance with a somewhat reduced flux of $(3.6\pm0.8)\times10^{-15}$~erg~s$^{-1}$~cm$^{-2}$~arcmin$^{-2}$ in 2--10~keV.
% A simple extrapolation of the spatial distribution found with observations mostly at low Galactic latitudes $\|b\| < 1$~deg would result in a flux similar to the CXB flux. 
% As we essentially need an empirical model for the sky background because we use the BGD and source regions simultaneously, we ignore the possible contribution of the Galactic Ridge X-ray Emission as long as we obtain an acceptable fit.

\section{Discussion and Conclusion} \label{sec-dis}
% \red{(UNDER REVISION)}
We have found extended X-ray emission around V4641 Sgr. The derived extension of $\sim 20$~pc assuming a Gaussian-like radial profile is smaller than that seen in gamma rays ($\approx 60$~pc with the HAWC observation: \citealt{hawc24_v4641}).
Here we first discuss the nature of the acceleration environment with an assumption that the X-rays are of non-thermal origin.
% \subsection{Morphology of the extended emission around V4641 Sgr}
% The X-ray morphology was found to be center-filled, which seems to be different from the gamma-ray morphology. 
% This is also different from the case of SS433, where a bipolar X-ray structure without significant emission around the center, similarly to its gamma-ray morphology, was reported \citep{kayama22, safi-harb22}.
% (\red{No eROSITA image published yet??}).
%
%\sout{As the acceleration sites are generally associated with bright X-ray emission,} %the center-filled X-ray morphology suggests that the acceleration sites should be close to V4641 Sgr ($\lesssim 10$~pc).
Although we cannot confidently identify the exact site of particle acceleration,
the X-ray radial profile describable with a Gaussian function centered at V4641 Sgr suggests that the acceleration site is close to the microquasar ($\lesssim 10$~pc).
% The primary origin of the gamma-ray emission is thus likely hadronic or electrons from another acceleration sites.
%
% Considering its very hard gamma-ray spectrum \citep{hawc24_v4641, lhaaso24_v4641}, the enhanced X-ray emission with relatively weak gamma-rays near V4641 Sgr may be consistent with the idea that high-energy electrons accelerated near V4641 Sgr diffuse around it while rapidly making the spectrum softer by losing highest-energy electrons.
The synchrotron cooling timescale for X-ray-emitting electrons is $t_{\rm cool} \approx 1000~{\rm yr} \, (E_{\rm e}/100~{\rm TeV})^{-1} (B/10~\mu{\rm G})^{-2}$, where the $E_{\rm e}$ and $B$ are the electron energy and magnetic field strength, respectively.
The diffusion length for this timescale is $R_{\rm dif} \approx (4 D(100~{\rm TeV})t_{\rm cool})^{0.5} \approx 80~{\rm pc}\, (D(100~{\rm TeV})/5\times10^{29}~{\rm cm}^2~{\rm s}^{-1})^{0.5} (E_{\rm e}/100~{\rm TeV})^{-0.5} (B/10~\mu{\rm G})^{-1}$, where the $D(E_{\rm e})$ is the diffusion coefficient. 
Thus, electrons with energies $\approx 100$~TeV can travel within the X-ray emitting region, if we assume the Galactic mean value for $D(E_{\rm e}) = 2\times10^{28}~{\rm cm}^2~{\rm s}^{-1}\,(E_{\rm e}/3~{\rm GeV})^{\delta}$ with $\delta = 0.3$--0.6 \citep{berezinskii90, ptuskin06}.
To match the observed X-ray extent $R_{\rm dif} \approx 10$~pc, either an enhanced magnetic field $B \approx 80~\mu$G with $D(E_{\rm e}) = 2\times10^{28}~{\rm cm}^2~{\rm s}^{-1}\,(E_{\rm e}/3~{\rm GeV})^{0.3}$ or a suppressed diffusion coefficient $D(100~{\rm TeV}) \approx 1\times10^{27}~{\rm cm}^2~{\rm s}^{-1}$ with an interstellar magnetic field level $B \approx 3~\mu$G is required.
%
% As these electrons cannot reach the gamma-ray peak locations as discussed in \cite{hawc24_v4641}, however, the primary origin of the gamma-ray emission is likely hadronic or electrons from another acceleration sites.

% \subsection{X-ray flux of the extended emission around V4641 Sgr}
% One important question about this system is the nature of its energetic gamma-ray emission. Information of the X-ray flux is of great importance in this aspect.
% % 
% If we assume a one-zone leptonic model to explain the emission from X-ray to PeV energies, the gamma-ray spectra by HAWC and LHAASO \citep{hawc24_v4641, lhaaso24_v4641} and the 2--10~keV flux $\approx 7\text{--}10 \times 10^{-12}$~erg~s$^{-1}$~cm$^{-2}$ would lead to a magnetic field strength $B \approx 3$--5~$\mu$G. We note that the X-ray flux assumed here is for the region covered by the Xtend FoV. So, if the X-ray flux for the entire gamma-ray emitting region is twice larger, the estimate will be $B \approx 4$--7~$\mu$G.
% Since these estimates are similar to the interstellar value without clue of amplification, the pure leptonic scenario will be unlikely\red{???}
%
% Contribution of the hadronic gamma-rays is claimed by \cite{hawc24_v4641, lhaaso24_v4641, neronov24}. Explaining both the X-ray and gamma-ray spectra with a one-zone hadronic model is challenging, however, because the observed X-ray flux is above the expected level of synchrotron emission from the secondary electrons, $< 10^{-12}$~erg~s$^{-1}$~cm$^{-2}$ \citep{neronov24}.
% , and the X-ray and gamma-ray morphologies are quite different.

\begin{figure}[htb!]
    \centering
    \includegraphics[width=8cm]{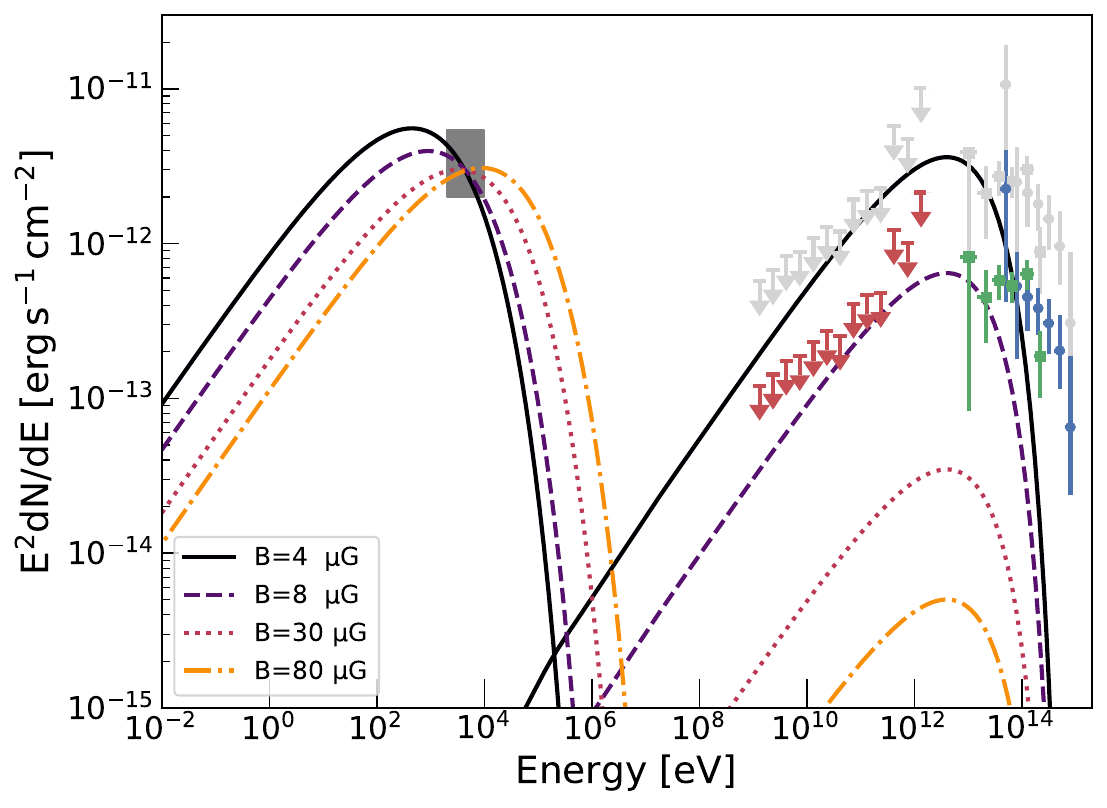}
    \caption{Broadband spectral energy distribution. The X-ray flux is derived for the XRISM source region (Figure~\ref{fig-image} (a)).
    The red, green, and blue flux points indicate gamma-ray spectra by Fermi-LAT, HAWC, and LHAASO \citep{neronov24,hawc24_v4641, lhaaso24_v4641}, respectively, scaled to the XRISM source region,
    while the gray points show spectra from the entire gamma-ray emitting region.
    Synchrotron and inverse Compton models with various magnetic field strengths are overlaid, assuming electrons with a power-law index of 2 and a cutoff energy of 50 TeV.
    The total energy content in electrons ($E>1$~TeV) is $2.6 \times 10^{47}$, $4.7 \times 10^{46}$, $2.5 \times 10^{45}$, and $3.6 \times 10^{44}$ erg for $B=$ 4, 8, 30, and 80~$\mu$G, respectively.
    }
    \label{fig-sed}
\end{figure}

Given that the very-high-energy gamma-ray emission is likely hadronic due to the very hard spectra with $\Gamma=2.2 \pm 0.2$ \citep{hawc24_v4641, lhaaso24_v4641}, one can constrain the lower limit of the magnetic-field strength based on the highest allowed level of the inverse Compton gamma-ray flux.
As shown in Figure~\ref{fig-sed}, we obtain $B \gtrsim 8~\mu$G, which indicates that the magnetic field is likely enhanced compared with the interstellar values.
%In Figure~\ref{fig-sed}, we assume a cutoff power-law model for electrons, with a spectral index of 2 and a cutoff energy of 50 TeV, and we examined various parameter sets, with the index of 2--3 and the cutoff energy of 10--100 TeV.
Since the spectral shape of accelerated electrons is completely unknown, we assume a cutoff power-law model and have examined various parameter sets, with the index of 2--3 and the cutoff energy of 10--100 TeV.
%Although the X-ray morphology suggests that electrons can fill the X-ray emitting region for $B \approx3$--80~$\mu$G before they get cooled,
Radiation cooling is also taken into consideration, as the X-ray spectrum is extracted from the region extending beyond $R_{\rm dif}$ (Figures~\ref{fig-image} and \ref{fig-radial}), where electrons may be affected by cooling. % for high magnetic field. % $B \gtrsim 10~\mu$G, where the cooling is not negligible.
In conclusion, the choice of the parameters and the cooling effect does not significantly affect the lower limit of $\approx 8~\mu$G.

If the X-ray emission is of thermal origin, one can discuss the physical process which produced the thermal plasma based on the electron temperature and plasma density. Assuming a shock heating as a representative scenario to generate X-ray emitting plasma in interstellar space, the derived electron temperature $\approx 3$~keV would require a shock velocity $\gtrsim 1500$~km~s$^{-1}$ (\citealt{xrism24_n132d, ohshiro24}). Plasma density $\sim 0.3$~ cm$^{-3}$ is typical for the interstellar medium.
% The total plasma mass is $\sim 27~M_{\odot}$. 
A supernova remnant would be able to explain these parameters, but the age estimate for the black hole in this system, $\gtrsim 10$~Myr \citep{salvesen20}, would contradict this scenario.
Thus, a shock originating from black hole activities, e.g., a jet termination shock, would be more feasible.
\red{If we assume a pair of jets along the line-of-sight direction \citep{hjellming00, orosz01}, strong and adiabatic shock, and homogeneous plasma,
the jet luminosity can be estimated from the shock kinetic energy as} $L_{\rm jet} \sim 2\times10^{39}~{\rm erg~s}^{-1}~ (n_{\rm ISM}/0.08~{\rm cm}^{-3}) (v_{\rm jet}/1500~{\rm km~s^{-1}})^3 (R/10~{\rm pc})^2$, where the $n_{\rm ISM}$, $v_{\rm jet}$, and $R$ are the interstellar plasma density, jet velocity, and radius of the extended emission, respectively. This is comparable to the Eddington luminosity of the system \citep{macdonald14, gandhi19}.
\red{This luminosity would not be unrealistic as the jet luminosity during the outburst in 1999 was inferred to exceed $10^{39}$~erg~s$^{-1}$ \citep{revnivtsev02}.
The spatial extent of the lobe produced by the jet to explain the observed extended X-rays, $\gtrsim 20$~pc, would be much larger than the typical size of the radio lobes found in microquasars such as Circinus~X-1, Cygnus~X-1, and GRS~1758$-$258 ($\lesssim 5$~pc; see a review by \citealt{gallo09}), but would be comparable to the X-ray extension of the SS~433/W50 complex ($\sim 30$~pc; \citealt{safi-harb22}). Such a large spatial extent may be related to the inferred high jet luminosity. \cite{hawc24_v4641} suggests that a jet with a luminosity of $10^{39}$~erg~s$^{-1}$ can, in principle, explain the extension of the TeV emission around V4641 Sgr ($\approx 60$~pc).}

% \red{Energetics of the jets.}
% The nearly spherical X-ray emission may prefer the supernova-remnant scenario. 

% \red{(Dust scattering)}
% The extended X-ray emission around V4641 Sgr is reminiscent of a dust scattering halo. 
An alternative origin to the extended X-ray emission around V4641 Sgr could be a dust scattering halo. 
The flux of the scattered photons can be evaluated based on the relation between an interstellar absorption column and optical depth in scattering \citep{predehl95}, the absorption column density $\approx 1.6\times10^{21}$~cm$^{-2}$, and a typical flux of V4641 Sgr during the bursting activity from September 2024, $\sim 10^{-11}\text{--}10^{-10}$~erg~s$^{-1}$~cm$^{-2}$ \footnote{See, e.g., the report on MAXI/GSC observations in September 2024 (\url{https://www.astronomerstelegram.org/?read=16804}) and XRISM results by Shidatsu et al. in preparation.}. The resultant flux level for the source region, $< 10^{-13}$~erg~s$^{-1}$~cm$^{-2}$ in 2--4~keV, is lower than the sky background. 
We have also examined 25-ks archival Chandra data of V4641 Sgr in 2002 (ObsID: 3800), \red{which were obtained during its quiescence ($\sim$five months after the outburst in May 2002)} with a low flux of V4641 Sgr $\sim 10^{-12}$~erg~s$^{-1}$~cm$^{-2}$ \footnote{Note that this observation covered a limited region where one cannot obtain a background region free from the extended source according to the extension measured with XRISM. Thus it is hard to detect the extended emission with this Chandra observation alone.}. With the same source and sky-background spectral models as the case (a) (Section~\ref{sec-spec}) plus a particle-background model for Chandra ACIS \citep{suzuki21b}, we obtained a surface brightness of the excess source, $\sim 7\times 10^{-15}$ erg~s$^{-1}$~cm$^{-2}$~arcmin$^{-2}$ in 2--10 keV, similar to that obtained with XRISM. This supports the idea that the extended emission is more persistent and thus independent of the bursting activity of V4641 Sgr.
\red{To conclude,} it is unlikely that the observed extended X-rays are due to dust scattering, \red{without specific conditions such as the presence of dense clumps in the line of sight}.

% \red{(UNDER REVISION)}
This work has discovered for the first time extended emission around V4641 Sgr with XRISM Xtend.
% If the X-ray and gamma-ray morphologies are confirmed to be different, 
The spatial extent of the X-ray emission $\sim 20$~pc in diameter is smaller than that seen in gamma rays. Assuming that the X-rays are of non-thermal origin, this suggest that the X-rays originate from an electron population near the acceleration sites, which is different from that responsible for the gamma-ray emission.
If the X-rays are of thermal origin, a jet termination shock would be able to explain the measured properties with a jet luminosity $\sim 2\times10^{39}$~erg~s$^{-1}$, which is comparable to the Eddington luminosity of this system.
Future radio observations and more extensive X-ray observations of the gamma-ray emitting region will help disentangle the nature of the acceleration environment.

% We note that, if the X-ray emission is mainly of thermal origin, the hadronic scenario will still be possible.

% In conclusion, we suggest that the extended X-ray emission originates from electrons at the acceleration sites close to V4641 Sgr, which may be different from the gamma-ray emitting particles.

\section*{Acknowledgment}
\red{We appreciate the helpful suggestions by the anonymous referee, which have improved the quality of this paper.} Part of this work was supported by JSPS KAKENHI grant numbers 22KJ3059, 24K17093 (H.S.), 
22K14064, 24H01819 (N.T.), 
24K17105 (Y.K.),
22K14028, 21H04487, 23H04899 (S.S.K.), 
19K14762, 23K03459, 24H01812 (M.S.),
% JSPS Core-to-Core Program (grant number: JPJSCCA20220002), 
NSF award 2209533,
the Tohoku Initiative for Fostering Global Researchers for Interdisciplinary Sciences (TI-FRIS) of MEXTs Strategic Professional Development Program for Young Researchers,
the Inter-University Research Programme of the Institute for Cosmic Ray Research (ICRR), University of Tokyo, Grant 2024i-F-05, Marco Perez Cisneros and the offices of CUCEI, Universidad de Guadalajara for the financial support of the research stays at ICRR in 2023 and 2024,
and
the Spanish \textit{Agencia estatal de investigaci'on} via PID2021-124879NB-I00.

% \facilities{XRISM (Xtend), Chandra (ACIS)}
% \software{HEAsoft \citep{heasarc2014},

\newpage
\bibliography{references}
\bibliographystyle{aasjournal}

\end{document}